\documentclass[twocolumn,preprintnumbers,amsmath,amssymb,superscriptaddress,footnote]{revtex4-2}
\usepackage{graphicx} 
\usepackage[dvipdfmx]{epsfig} 
\usepackage{bm}
\usepackage{ulem}
\usepackage{amsmath,color}

\def\m@thcombine#1#2{%
  \setbox0=\hbox{$#1$}
  \setbox1=\hbox{$#2$}
  \ifdim\wd0>\wd1
    \setbox0=\hbox to\wd1{\hss\box0\hss}
  \else
    \setbox1=\hbox to\wd0{\hss\box1\hss}
  \fi
  \mathop{\vcenter{
    \offinterlineskip\box0\box1}}}
\def\lesim{\m@thcombine<\sim}
\def\gesim{\m@thcombine>\sim}

\begin{document}
\title{Large enhancement of total reaction cross sections
  at the edge of the island of inversion in Ti, Cr, and Fe isotopes}
\author{W. Horiuchi}
\email{whoriuchi@nucl.sci.hokudai.ac.jp}
\affiliation{Department of Physics,
  Hokkaido University, Sapporo 060-0810, Japan}

\author{T. Inakura}
\affiliation{Laboratory for Zero-Carbon Energy, Institute of Innovative Research, Tokyo Institute of Technology, Tokyo 152-8550, Japan}

\author{S. Michimasa}
\affiliation{Center for Nuclear Study, The University of Tokyo, 2-1 Hirosawa, Wako, Saitama 351-0198, Japan}

\begin{abstract}
A systematic analysis of nuclear deformation
is made for neutron-rich Ti, Cr, and Fe isotopes
to explore the nuclear structure in the island of inversion near $N=40$,
where strong nuclear deformation is predicted.
The nuclear ground states are obtained by the Skyrme Hartree-Fock
method in three-dimensional coordinate space,
which properly describes any nuclear shape.
Three types of Skyrme interactions are employed
to generate various deformed states in its isotopic chain.
We find that in the island of inversion
the occupation of highly elongated intruder orbits
induces not only large quadrupole deformation
but also large hexadecapole deformation.
This appears as a sizable enhancement of the nuclear matter radius,
showing the characteristic shell effect
of the density profile near the nuclear surface.
We show that the edge of the island of inversion,
where the intruder orbit starts being occupied,
can be determined by measuring the enhancement
of the total reaction cross section at high incident energy.
The possibility of constraining the hexadecapole deformation
by a measurement of the total reaction cross sections is discussed.
\end{abstract}
\maketitle

\section{Introduction}

Recent theoretical and experimental studies on short-lived nuclei 
are addressing the nature of a proposed island of inversion region near $N=40$, 
and suggest that nuclear stability enhances in neutron-rich isotones 
around $^{64}$Cr~\cite{Lenzi10} and $^{62}$Ti~\cite{Michimasa20}.
A concept of the island of inversion~\cite{Warburton90} was first proposed 
to explain unexpected nature 
such as excess stabilization of atomic masses and low-lying first $2^+$ energies of even-even nuclei 
around $^{32}$Mg~\cite{Thibault75,Detraz79,Motobayashi95}.
Similar scenarios were suggested to explain the onset of nuclear deformation in neutron-rich magic nuclei 
at $N=8$~\cite{Navin00,Iwasaki00a,Iwasaki00b} and 
$28$~\cite{Glasmacher97,Sarazin00,Bastin07,Takeuchi12}.
Also, the Jahn-Teller stabilization at $N=40$ was theoretically predicted~\cite{Brown01}.
Since this suggestion, many experimental results
on the $N=40$ island of inversion 
were reported: The systematics of
the atomic masses~\cite{Michimasa20,Naimi12,Meisel16,Mougeot18},
low-lying excited states~\cite{Hannawald99,Sorlin03,Adrich08,Aoi09,Gade14,Wimmer19,Cortes20}, and quadrupole collectivity~\cite{Liungvall10,Crawford13}.
Despite the above experimental information in this mass region, 
the whole picture of the island of inversion near $N=40$
is still uncertain such as divergence between the peaks of the mass
stability and the quadrupole collectivity. 
Therefore, careful investigations through various observables
are crucially important.

Strong nuclear deformation is
one of the most prominent indications of the island of inversion.
In this paper, the edge of the island of inversion near $N=40$
is defined as the occupation of the intruder orbit
stemming from the spherical $0g_{9/2}$ orbit, leading to large deformation,
which is a natural extension of the island of inversion near $N=20$,
where the intruder orbit from the spherical $0f_{7/2}$ orbit is occupied.
We remark that the effect of the configuration mixing around $^{64}$Cr was
discussed in detail~\cite{Lenzi10}.
To know the structure information on the occupation of the intruder orbits
related to $fp$- and $gds$-configuration mixing,
it is more advantageous to study observables such as
transition probabilities and nuclear radii.

Generally, the direct determination of the nuclear deformation
has some difficulty, e.g., one has to assume
a simple structure model to extract the quadrupole
deformation parameter from observed electric quadrupole transition
strength~\cite{Pritychenko16}.
Meanwhile, one promising measure that reflects the nuclear deformation
is the nuclear matter radius.
The nuclear deformation drastically changes
the density profiles near the nuclear surface,
leading to the enhancement of the nuclear radius.
Measuring the total reaction or interaction cross section 
at high incident energy has been one of the standard methods
and found various exotic phenomena such as halos~\cite{Tanihata85,Tanihata13},
developed neutron skin~\cite{Suzuki95}, in neutron-rich nuclei far from the $\beta$ stability line.
Recent developments of the radioactive beam facility
extend the applicable mass region:
Near dripline nucleus $^{29}$F~\cite{Bagchi20}
and medium-mass nuclei beyond $N=28$, $^{42-51}$Ca~\cite{Tanaka20}.
One of the advantages of the total reaction cross section study is
that the reaction theory has been well tested,
allowing us to directly relate the cross section with the nuclear radius.
Extracting nuclear size properties in the isotopic chain reveals
structure changes due to excess neutrons.
A systematic measurement of the cross sections for neutron-rich Ne and Mg
clearly showed the evolution of the nuclear deformation
with the help of reliable microscopic theoretical models~\cite{Takechi12,Minomo11,Minomo12,Sumi12,Horiuchi12,Watanabe14,Takechi14,Horiuchi15}.
The sudden increases of the total reaction cross section
at the edge of the island of inversion near $N=20$ were explained 
by the enhancement of the nuclear matter radius
coming from a diffused nuclear surface induced
by the strongly deformed nuclear state.

Extending this idea, in this paper, we study the nuclear shape
of Ti, Cr, and Fe isotopes
and discuss the possibility to determine the edge of
the island of inversion near $N=40$ using the total reaction cross sections.
  The territory of the island of inversion
  in this mass region has been explored in large-scale shell-model
  calculations~\cite{Lenzi10}.
  The experimental indication that the border of the
  island of inversion goes beyond $N=50$
  was obtained in Ref.~\cite{Santamaria15}.
  This work will show the utility of the total reaction cross section
  measurement to determine the location
of the island of inversion in the nuclear landscape.
In the Ti, Cr, and Fe isotopes,
the nuclear deformation is strongly model dependent in $N = 34$--$42$.
Though most of the nuclei have a prolate shape, which is the so-called
prolate dominance~\cite{Kumar70,Tajima01,Hamamoto09},
some nuclei may exhibit oblate deformation.
The shape of the wave function is determined with a delicate balance
of the single-particle (s.p.) energies as the energy surface
is soft with respect to the quadrupole deformation
parameter $\gamma$ in this mass region~\cite{Yoshida11,Sato12}.
Recently, the deformation effect on the nuclear density profile
was discussed in detail~\cite{Horiuchi21}.
The nuclear density was changed
not only for the surface region but also the internal region
by the nuclear deformation.
Here we systematically investigate how
those different density profiles are observed
in the total reaction cross sections of Ti, Cr, and Fe isotopes
at high incident energy.

The paper is organized as follows. In the next section, we briefly describe
the microscopic structure and reaction models employed in this paper.
Setups of the Skyrme-Hartree-Fock (HF) method and the Glauber model
are given in Secs.~\ref{HF.sec} and \ref{Glauber.sec}, respectively.
We perform the HF calculations
in the three-dimensional Cartesian coordinate,
which can express any deformed shape.
The model dependence is investigated by examining
three sets of the Skyrme-type effective interactions.
Using the density distributions obtained by the HF calculations,
we compute the total reaction cross sections using the Glauber model
without introducing any adjustable parameters.
Section~\ref{Results.sec} presents our results.
First, in Sec.~\ref{def.sec} we discuss structure changes
of Ti, Cr, and Fe isotopes near $N=40$,
especially focusing on their nuclear deformation
at around the neutron numbers $N=34$, 36, and 40
by the microscopic mean-field model.
A comparison of the calculated results with the experimental evaluation
including recent data of the two-neutron separation energy is made.
Following this comparison, we show in Sec.~\ref{RCS.sec}
a systematic measurement of the total reaction cross sections
becomes important
to determine the location of the island of inversion near $N=40$.
Section~\ref{hex.sec} discusses the role of the intruder s.p. orbits
to determine the nuclear shape, especially
focusing on the hexadecapole deformation.
In Sec.~\ref{diff.sec},
we address the possibility to determine these deformation parameters
from the nuclear radius using a macroscopic model approach.
The characteristics of the density profile
of the nuclei in the island of inversion are elucidated.
Conclusion is made in Sec.~\ref{Conclusions.sec}.

\section{Methods}
\label{Methods.sec}

\subsection{Skyrme Hartree-Fock calculation in three-dimensional coordinate
  space}
\label{HF.sec}

In this paper, we employ the Skyrme-HF calculation
in three-dimensional (3D) coordinate representation.
Since all details can be found in Refs.~\cite{Inakura06, Horiuchi12,Horiuchi20},
we only give a minimum explanation for the present analysis.
The ground-state wave function is expressed as
the product of deformable s.p. orbits
represented by the 3D Cartesian mesh
which is flexible enough to describe higher-order multipole deformation,
such as the hexadecapole one.
We obtain these s.p. orbits fully self-consistently
in the sphere of radius 20 fm
based on the Skyrme energy density functional~\cite{Vautherin72},
in which the total energy is
a functional of the intrinsic density $E[\rho_{\rm int}]$
and is minimized using the imaginary-time method~\cite{Davies80}.
Three kinds of Skyrme parameter sets, 
SkM$^\ast$~\cite{SkMs}, SLy4~\cite{SLy4} and SkI3~\cite{SkI3}
are employed to obtain various density profiles.
The SkM$^\ast$ interaction is one of the most used Skyrme interactions
for nuclear structure calculation.
The SLy4 interaction is constructed to reproduce a theoretical equation
of state~\cite{Akmal98} and 
experimental data in a wide mass region, especially for neutron-rich nuclei.
The SkI3 interaction is
designed with attention to the reproducibility of s.p. levels of $^{208}$Pb.
  There have been proposed
  a lot of the Skyrme interaction sets in the market~\cite{Dutra12}.
  Investigations of these three interactions are useful as they
  produce, e.g., different deformations and isovector density profiles~\cite{Horiuchi12,Horiuchi17,Horiuchi21}.
  The pairing interaction may change
  the nuclear deformation and induces the fractional occupation probability
  near the Fermi level.
  We remark that an elaborated beyond-mean-field calculation
  was done in this mass region~\cite{Tomas16}.
  The purpose of this study is
  to elucidate the effect of various nuclear deformations on the total reaction
  cross section. For the sake of simplicity,
  we only include the nuclear deformation, which is the most essential
  ingredient to determine the nuclear density profile.
  The pairing correlation is ignored in the present analysis
  as it induces further model dependence~\cite{Stoitsov03,Delaroche10,Changizi15,Horiuchi16,Horiuchi17}.

Once the ground-state wave function is obtained,
the mean value of an operator $X(x,y,z)$ can be obtained by
\begin{align}
  \left<X\right>=\frac{1}{A}\int_{-\infty}^\infty dx\int_{-\infty}^\infty dy\int_{-\infty}^\infty dz\,\rho_{\rm int}(x,y,z)X(x,y,z).
\end{align}
The quadrupole deformation parameter is calculated as 
\begin{align}
  \beta_2&=\sqrt{\beta_{20}^2+\beta_{22}^2},
\label{beta2.eq}
\end{align}
where
\begin{align}
  \beta_{20}&=\sqrt\frac{\pi}{5}\frac{\left<2z^2-x^2-y^2\right>}{\left<r^2\right>},\\
  \beta_{22}&=\sqrt\frac{3\pi}{5}\frac{\left<x^2-y^2\right>}{\left<r^2\right>},
\end{align}
with $r^2=x^2+y^2+z^2$. We take $z$ as the quantization axis 
and choose it as the largest (smallest) principal axis for
prolate (oblate) deformation.
  We note that the ground-state wave function can be triaxially deformed
  as indicated by
  $\tan \gamma=\beta_{22}/\beta_{20}$ with $0^\circ <\gamma < 60^\circ$
  in the present calculations.
 
The ground-state wave function may exhibit higher-order multipole deformation.
We also calculate the hexadecapole deformation parameter
defined by, e.g., see Ref.~\cite{Bender03},
\begin{align}
  \beta_4 =\left[\sum^4_{m=-4} \beta^2_{4m}\right]^{\frac{1}{2}}
  \label{beta4.eq}
\end{align}
with
\begin{align}
\beta_{4m} = \frac{4\pi}{3R^4} \langle r^4 Y_{4m} \rangle,
\end{align}
where $R=\sqrt{\frac{5}{3}\left<r^2\right>}$ is the nuclear radius.
For later convenience, we define the hexadecapole moment
operator as $Q_{4m}=r^4Y_{4m}$, more explicitly with $m=0$
\begin{align}
  Q_{40}=\frac{3}{16}\sqrt{\frac{1}{\pi}}(35z^4-30z^2r^2+3r^4).  
\label{q40.eq}
\end{align}

\subsection{Total reaction cross sections by Glauber theory}
\label{Glauber.sec}
  
To bridge a gap between the density profiles
obtained by the structure calculation
and reaction observables, we need an appropriate reaction model.
Here we consider the total reaction cross section on a carbon target
at a high incident energy of more than a hundred MeV.
The Glauber theory~\cite{Glauber} formulated based
on the adiabatic and eikonal approximations efficiently
describes high energy nucleus-nucleus collisions of interest.
In the Glauber formalism,
the total reaction cross section is evaluated by
\begin{align}
\sigma_R=\int\,d\bm{b} \left(1-\left|e^{i\chi(\bm{b})}\right|^2\right),
\end{align}
where the squared modulus of the phase-shift function
$e^{i\chi(\bm{b})}$ is integrated over the impact parameter
vector $\bm{b}$. The evaluation of $e^{i\chi(\bm{b})}$ is in general
demanding because it involves multi-dimensional
integration~\cite{Varga02,Nagahisa18,Hatakeyama19}.
To incorporate the multiple scattering effect efficiently,
here we employ the nucleon-target formalism (NTG) as given in Ref.~\cite{NTG}:
\begin{align}
  &i\chi(\bm{b})\approx
  -\int d\bm{r}^P\,\rho^P(\bm{r}^P)\notag\\
  &\times\left[1-\exp\left\{-\int d\bm{r}^T\rho^T(\bm{r}^T)\Gamma_{NN}(\bm{s}^P-\bm{s}^T+\bm{b})\right\}\right]
\end{align}
where $\bm{r}^{P(T)}=(\bm{s}^{P(T)},z^{P(T)})$
denotes the two dimensional coordinate
of the projectile (target) nucleus perpendicular
to the beam direction, $z^{P(T)}$.
The NTG includes higher multiple-scattering terms
and is known to give a better description than optical-limit approximation~\cite{Horiuchi06, Horiuchi07, Ibrahim09, Nagahisa18}; hence
it has been employed as a standard tool to analyze
the nuclear matter radius from measured cross sections~\cite{Takechi09, Kanungo10, Kanungo11, Takechi12, Takechi14, Bagchi20}.
The theory requires the projectile
$\rho^P(\bm{r}^{P})$ and target density distributions $\rho^T(\bm{r}^T)$
and the profile function $\Gamma_{NN}$.
We employ for the target nucleus $^{12}$C
the harmonic-oscillator type density~\cite{Ibrahim09}
that reproduces the rms point-proton radius
of $^{12}$C, 2.33 fm~\cite{Angeli13}.
The parameters of the profile function are taken from Ref.~\cite{Ibrahim08},
which has been well tested, showing satisfactory descriptions
of nucleus-nucleus collisions including short-lived nuclei, e.g.,
in Refs.~\cite{Horiuchi10, Horiuchi12, Horiuchi15, Horiuchi16, Nagahisa18}.
We use the density distributions obtained from
the HF calculations as the input projectile density distributions
obtained by averaging over angles $\hat{\bm{r}}=(\theta,\phi)$~\cite{Horiuchi12}
\begin{align}
\rho^{P}(r)=\frac{1}{4\pi}\int d\hat{\bm{r}}\,\rho_{\rm int}(r,\hat{\bm{r}}).
\end{align}
Note that in this work we treat all the physical quantities
   in the intrinsic frame, e.g., without angular momentum projection.
   The validity of this averaging treatment of the intrinsic density
   was confirmed in Ref.~\cite{Sumi12} through a comparison
   of the angular-momentum-projected density.
Since the theory has no adjustable parameter,
the total reaction cross section properly reflects
the characteristics of the density profile
obtained from the microscopic structure model.

\section{Results and discussions}
\label{Results.sec}

\subsection{Nuclear quadrupole deformation and structure of Ti, Cr, and Fe isotopes}
\label{def.sec}

\begin{figure*}[ht]
  \begin{center}
    \epsfig{file=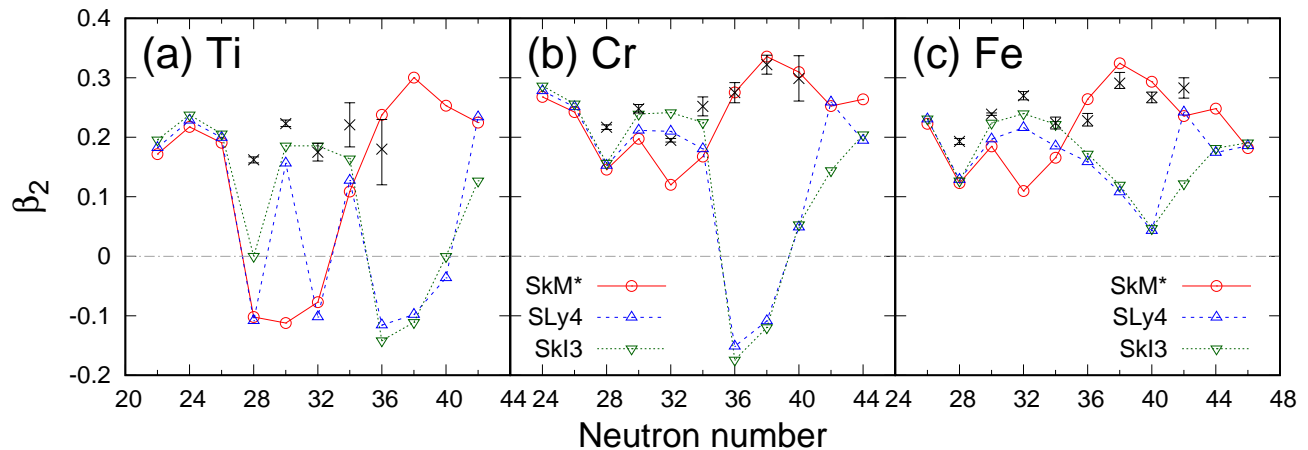, scale=1.3}
    \caption{Quadrupole deformation parameter $\beta_2$ of
      (a) Ti, (b) Cr, and (c) Fe isotopes as a function of the neutron number.
Triaxial deformation
      with $0^\circ<\gamma < 30^\circ$ ($30^\circ <\gamma < 60^\circ$)
      is treated as prolate (oblate).
      The thin lines that connect theoretical results are to guide the eye.
      The cross symbols with an error bar indicate
      the empirical $|\beta_2|$ values evaluated by measured
      quadrupole transition strength~\cite{Pritychenko16}.
    }
    \label{beta2.fig}
  \end{center}
\end{figure*}

Figure~\ref{beta2.fig} displays the calculated
quadrupole deformation parameter of Ti, Cr, and Fe isotopes.
For convenience,
we use the familiar notation $-\beta_2$ for oblate shapes.
As we see in the figure,
the nuclear deformation is strongly interaction dependent
in $N=34$--42 because the energy surface is soft with respect
to the quadrupole deformation parameter $\gamma$
in this mass region~\cite{Yoshida11,Sato12} like as near $N=20$.
In fact, some nuclear states exhibit a triaxial shape:
$^{62}$Ti ($\gamma=8.1^\circ$) for SkM$^*$;
$^{56}$Cr ($\gamma=14.3^\circ$),
$^{60}$Cr ($\gamma=32.5^\circ$), and
$^{64}$Fe ($\gamma=28.2^\circ$) for SLy4;
and
$^{54}$Ti ($\gamma=24.4^\circ$),
$^{58}$Ti ($\gamma=43.5^\circ$),
$^{60}$Cr ($\gamma=39.2^\circ$),
$^{58}$Fe ($\gamma=17.2^\circ$),
$^{62}$Fe ($\gamma=19.6^\circ$), and
$^{64}$Fe ($\gamma=21.9^\circ$) for SkI3.
Since the effect of the triaxiality
is small for the enhancement of the nuclear radius
as verified for Ne isotopes~\cite{Sumi12}, quadrupole deformation
with $0^\circ<\gamma < 30^\circ$ ($30^\circ <\gamma < 60^\circ$)
is treated as prolate (oblate) in the figure for simplicity.   
The SkM$^*$ interaction tends to give strong deformation
and the other interactions favor a less deformed shape.
The most striking difference appears at $N \geq 34$
in which the prolate deformation grows for the SkM$^*$ interaction,
while the others exhibit a less deformed shape.
These differences can be attributed to the neutron
s.p. level structure which can be explained in a similar way
like in the island of inversion found near $N=20$.
On moving $N=34$ to 36, the SkM$^*$ interaction makes the occupancy
of the intruder orbit with the asymptotic quantum number
$[nn_z\Lambda]\Omega=[440]{1/2}$ ~\cite{Nilsson55} stemming from
the spherical $0g_{9/2}$ orbit and causes large prolate deformation.
On the other hand, the SLy4 and SkI3 interactions increase
the occupancy of the $fp$-shell orbit that results
in much smaller $\beta_2$ values compared with the SkM$^*$ interaction.
This is because the $0f_{5/2}$-$0g_{9/2}$ level spacings
for the SLy4 and SkI3 interactions are larger than that for
the SkM$^*$ interaction. In fact, the calculated
$0f_{5/2}$-$0g_{9/2}$ level spacings in a spherical $^{90}$Zr
are 3.14 MeV, 4.99 MeV, and 7.09 MeV for the SkM$^*$, SLy4,
and SkI3 interactions, respectively.
For the SLy4 and SkI3 interactions, since the $0f_{5/2}$--$0g_{9/2}$
level spacings are large,
the occupation of the $[440]1/2$ orbit, i.e.,
two-particle-two-hole (2p-2h) state,
requires more energy than the energy that can be obtained from reduction,
whereas the occupation of $[440]1/2$ orbit
is realized with the SkM* interaction
because of the small $0f_{5/2}$-$0g_{9/2}$ level spacing.
With the SkM$^*$ interaction,
the largest deformation for Ti, Cr, and Fe isotopes at $N = 38$
is found with occupancy of the $[440]{1/2}$ and $[431]{3/2}$ orbits.
At $N=42$, the SLy4 interaction prefers additional occupancy
of the orbit originated from the spherical $0g_{9/2}$ orbit,
yielding different $N$-dependence of the nuclear deformation from
the SkI3 interaction.

Since the nuclear structure is strongly depends
on the Skyrme interaction employed,
it is needed to verify these theoretical models
through a comparison with experimental data.
Figure \ref{beta2.fig} also displays the experimental evaluations
of $|\beta_2|$~\cite{Pritychenko16}. 
  Though they are model dependent,
  relying on a simple collective model and
  only show the magnitude of the quadrupole deformation,
  their trend can be a guide to the nuclear deformation
  in this mass region.
All the evaluated $|\beta_2|$ values exhibit large
quadrupole deformation $N\geq 36$.
Their trends in Cr and Fe isotopes follow the results of
the SkM$^*$ interaction.  
The SLy4 and SkI3 interactions are relatively less reproductive,
predicting small collectivity for the Cr and Fe isotopes at $N = 36$--40.
In Ti isotopes, the magnitudes of the experimental evaluations are 
consistent with all the theoretical results
because of their large uncertainties. 
Based on the systematical relation between the $|\beta_2|$ value
and $2_1^+$ energies, the low-lying $2_1^+$ states in $^{60,62}$Ti strongly
show large quadrupole collectivity~\cite{Gade14,Cortes20}.
Therefore, also in Ti isotopes, the SkM$^*$ interaction
appears to be the most reasonable among the three interactions
employed in this paper.

\begin{figure}[ht]
  \begin{center}
    \epsfig{file=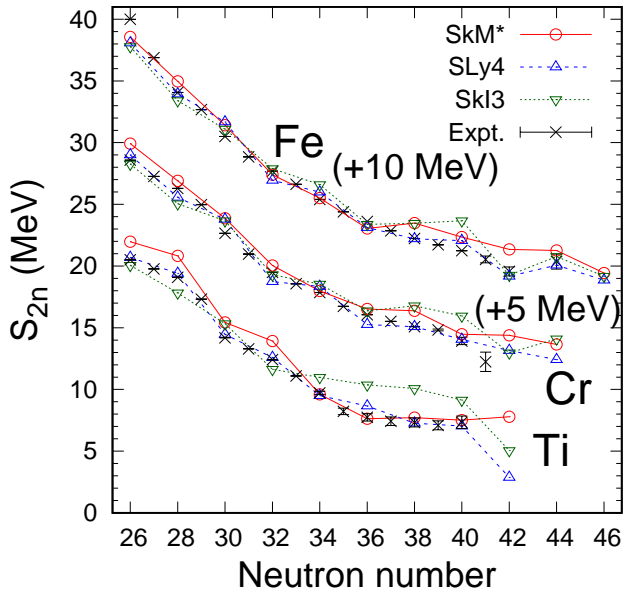, scale=1.35}
    \caption{Two-neutron separation energies
      of Ti, Cr, and Fe isotopes as a function of the neutron number.
      For clarity, 5 MeV for Cr and 10 MeV for Fe are
      respectively added to the results.
      The SkM$^*$, SLy4, and SkI3 interactions are employed.
      The theoretical results for $N$ even nuclei are plotted.
      The lines are to guide the eye.
      The experimental data are taken
      from Refs~\cite{Michimasa20,Meisel20,AME2020}.
    }
    \label{s2n.fig}
  \end{center}
\end{figure}

Since the nuclear deformation is sensitive to the energy levels
near the Fermi level, we also compare our results
with available experimental data of the two-neutron separation
energy, which has intensively been studied in recent years. 
Figure~\ref{s2n.fig} plots the two-neutron separation
energies ($S_{\rm 2n}$) of Ti, Cr, and Fe isotopes.
The experimental data is taken from the AME2020 database~\cite{AME2020} 
and recent measurements~\cite{Michimasa20,Meisel20}.
Overall agreement is obtained for all interactions employed here
around $N = 34$--$36$.
In $N \gtrsim 34$, the SkI3 interaction tends to overestimate the data. 
The SLy4 interaction gives excellent agreement with the experimental data.
The SkM$^*$ prediction nicely follows the experimental results in Ti isotopes
but slightly overestimates the data in Cr and Fe isotopes.
It appears that these effective interactions
may not be accurate enough to describe
the two-neutron separation energies in this mass region.
We note, however, that the behavior does not follow
that of the nuclear deformation
though we see notable differences in the nuclear deformations
as given in Figs.~\ref{beta2.fig} and \ref{beta4.fig}.
On the other side,  the $S_{\rm 2n}$ trends may indicate that 
the enhancement of nuclear stability in this region could be affected 
not only from the quadrupole deformation 
but also from the other effects, e.g.,
core swelling effect~\cite{Horiuchi20, Horiuchi21}.

\subsection{Nuclear radii and total reaction cross sections in the island of inversion near $N=40$}
\label{RCS.sec}

\begin{figure}[ht]
\begin{center}
  \epsfig{file=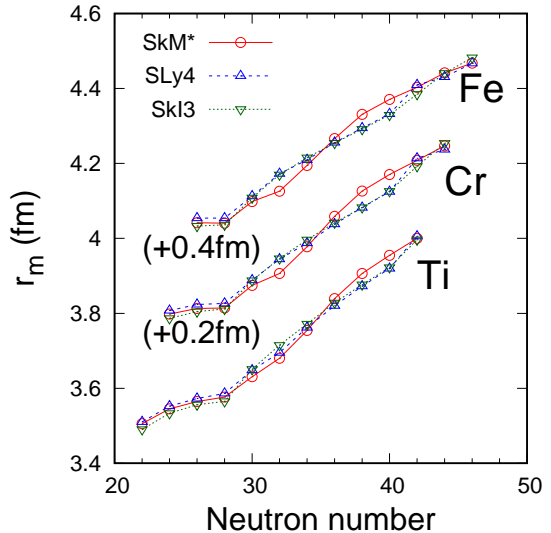, scale=1.2}                    
    \caption{Rms matter radii of Ti, Cr, and Fe isotopes
      as a function of the neutron number.
      For clarity, 0.2 and 0.4 fm are added
      to the results of Cr and Fe, respectively.
      The lines are to guide the eye.     
      The SkM$^*$, SLy4, and SkI3 interactions are employed.}
    \label{radii.fig}
  \end{center}
\end{figure}

\begin{figure*}[ht]
  \begin{center}
    \epsfig{file=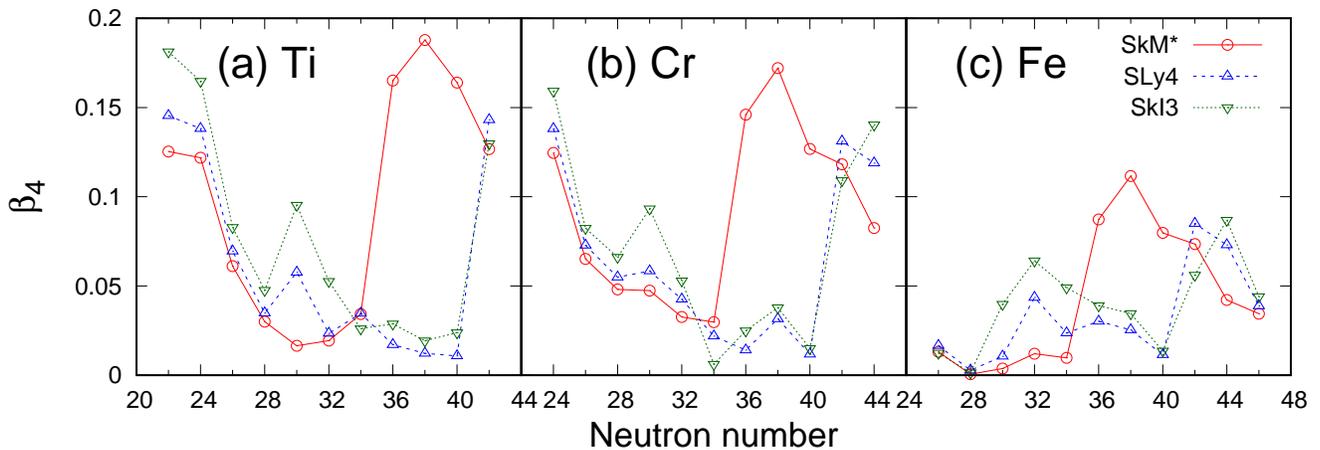, scale=1.3}        
    \caption{Hexadecapole $\beta_4$
      deformation parameters of
      (a) Ti, (b) Cr, and (c) Fe isotopes as a function of the neutron number.
    }
    \label{beta4.fig}
  \end{center}
\end{figure*}

In Sec.~\ref{def.sec}, we overviewed the experimental situation
on the structure of the Ti, Cr, and Fe isotopes and found
that more experimental data is needed to establish 
the structure of those isotopes near $N=40$.
This motivates us to study the systematics of nuclear matter radii,
which is sensitive to changes in the nuclear deformation.
Figure~\ref{radii.fig} displays the root-mean-square (rms) matter radii
of Ti, Cr, and Fe isotopes, $r_{\rm m}=\sqrt{\left<r^2\right>}$.
For the same Skyrme interaction,
similar neutron-number dependence
is predicted for the rms radii of those isotopes.
However, the SkM$^{*}$ interaction exhibits
different characteristics compared with the SLy4 and SkI3 ones.
The SkM$^{*}$ predicts a sudden increase of the rms radius at $N = 34$, 
which is exactly a consequence of the onset of the intruder configuration
and resultant nuclear deformation 
as given in Figs.~\ref{beta2.fig} and ~\ref{beta4.fig}.
The nuclear radii obtained with the SLy4 and SkI3 interactions
are largely enhanced not at $N = 34$ but at $N=42$, 
which are comparable to those obtained with the SkM* interaction.
These calculations clearly demonstrate a strong correlation
between the occupation of the intruder configuration
and the sudden increase of the nuclear radius.
Therefore the edge of the island inversion
can be observed experimentally
as the sudden increase of the nuclear radius.

\begin{figure}[ht]
  \begin{center}
    \epsfig{file=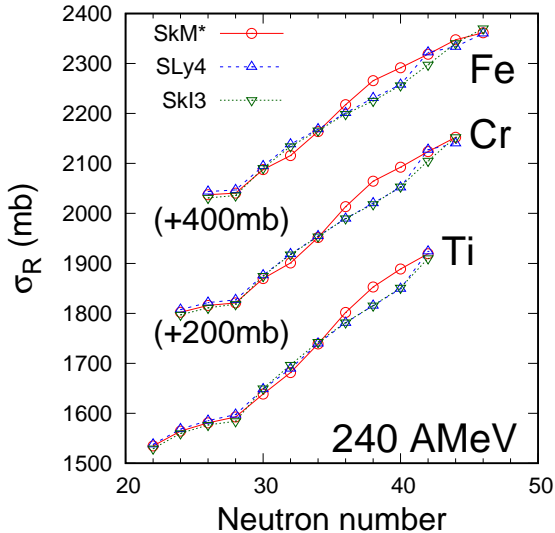, scale=1.2}                        
    \caption{
      Total reaction cross sections on a carbon target
      at 240 MeV/nucleon of Ti, Cr, and Fe isotopes
      as a function of the neutron number.
      For clarity, 200 and 400 mb are added
      to the results of Cr and Fe, respectively.
      The lines are to guide the eye.      
      The SkM$^*$, SLy4, and SkI3 interactions are employed.}
    \label{sigR.fig}
  \end{center}
\end{figure}

We examine how those differences are significant
in cross section measurement.
It is again stressed that the cross section calculations do not include
adjustable parameters and the reliability of the adopted Glauber model
has been established.
As inputs to the theory are the density distributions
obtained by the microscopic mean-field model,
a systematic trend of the cross sections will properly describe
the structure changes owing to the nuclear deformation.
Figure~\ref{sigR.fig} displays the total reaction cross sections on a carbon
target. The incident energy is chosen as 240 MeV/nucleon,
where the recent interaction
cross section measurements were performed~\cite{Tanaka20,Bagchi20}.
The differences of the nuclear radii in $34<N<40$ are further emphasized in
the total reaction cross sections using a carbon target
because the carbon target has more sensitivity
of the density profile beyond the nuclear surface~\cite{Horiuchi12,Horiuchi14}.
The cross section differences in $34<N<40$ 
are large at most more than $2$\%, which can be distinguished by measurement
as its uncertainty is typically $\lesssim 1$\%~\cite{Tanaka20,Bagchi20}.
A systematic cross section measurement of these isotopes
is of cardinal importance because it will offer further evidence
to determine the edge of the island of inversion near $N=40$,
where the strong deformation is predicted.

\subsection{Enhancement of hexadecapole deformation in the island of inversion}
\label{hex.sec}

In this section, we discuss a unique feature of
the nuclear deformation in the island of inversion.
As discussed in Sec.~\ref{def.sec},
the occupation of the $[440]1/2$ orbital is a key to
determine the quadrupole deformation in $N\gtrsim 34$.
Because this is the most elongated along $z$ axis,
the occupancy may induce higher multipole deformation, i.e.,
hexadecapole deformation.
Figure~\ref{beta4.fig} plots the hexadecapole deformation
parameter $\beta_4$ for Ti, Cr, and Fe isotopes. As we see in the figure,
the $\beta_4$ value drastically increases
at $N=36$ for SkM* and $N=42$ for SLy4 and SkI3.
where the $[440]1/2$ orbit is occupied, i.e., the island inside.

\begin{figure}[ht]
  \begin{center}
    \epsfig{file=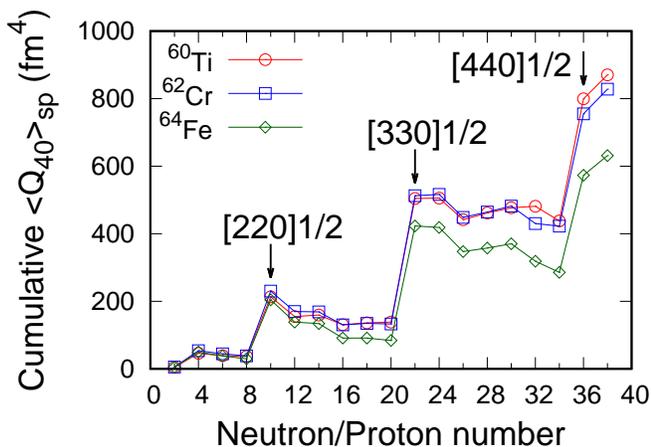, scale=1.4}
    \caption{Cumulative single-particle
      hexadecapole moment of the $N=38$ isotones.
      The arrows indicate $N,Z=10, 22$, and $N=36$,
      where $[nn0]1/2$ orbit is occupied.
      The SkM$^*$ interaction is employed.
    See text for details.}
    \label{cumulative.fig}
  \end{center}
\end{figure}

\begin{figure*}[htb]
  \begin{center}
    \epsfig{file=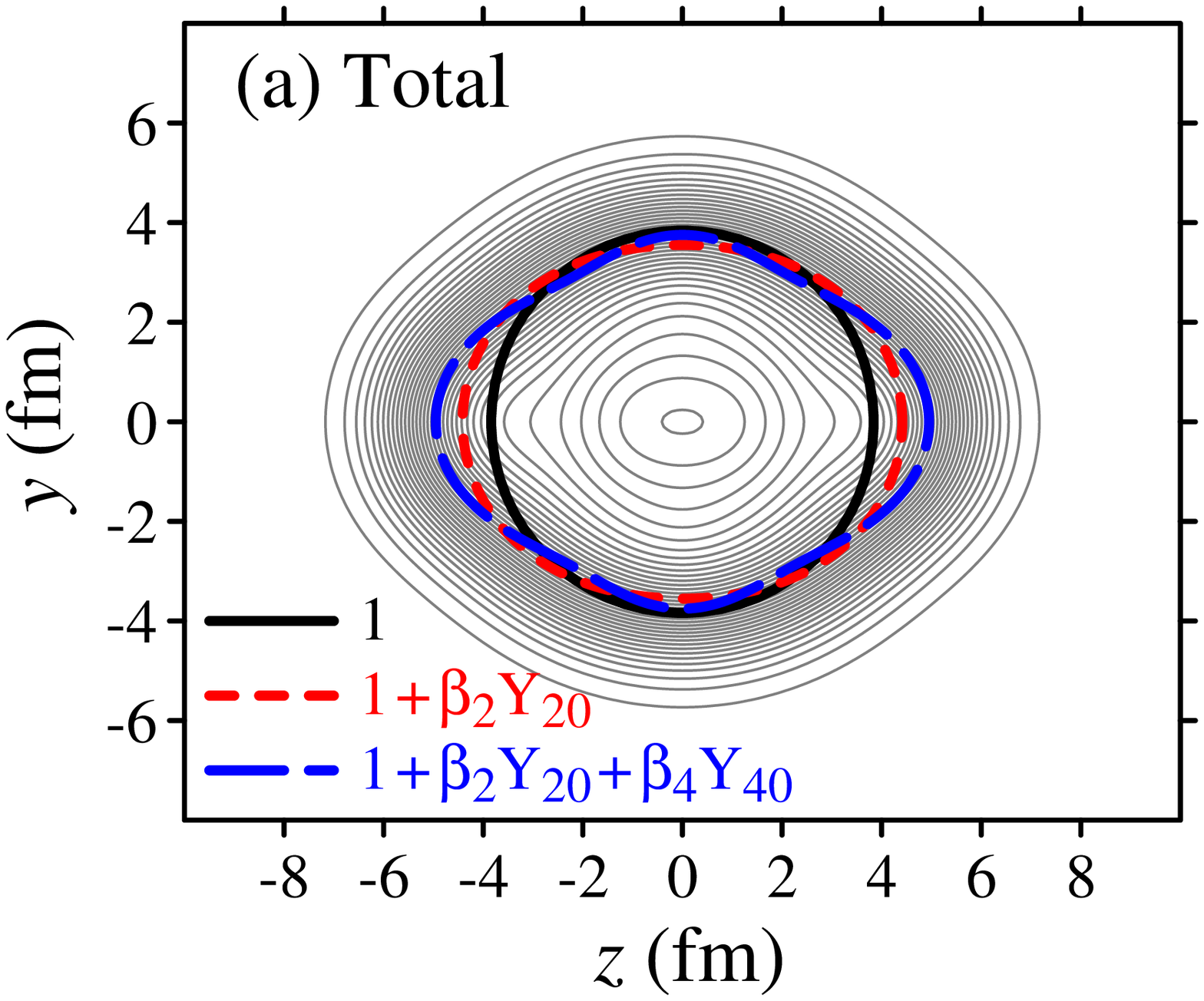,scale=0.31}
    \epsfig{file=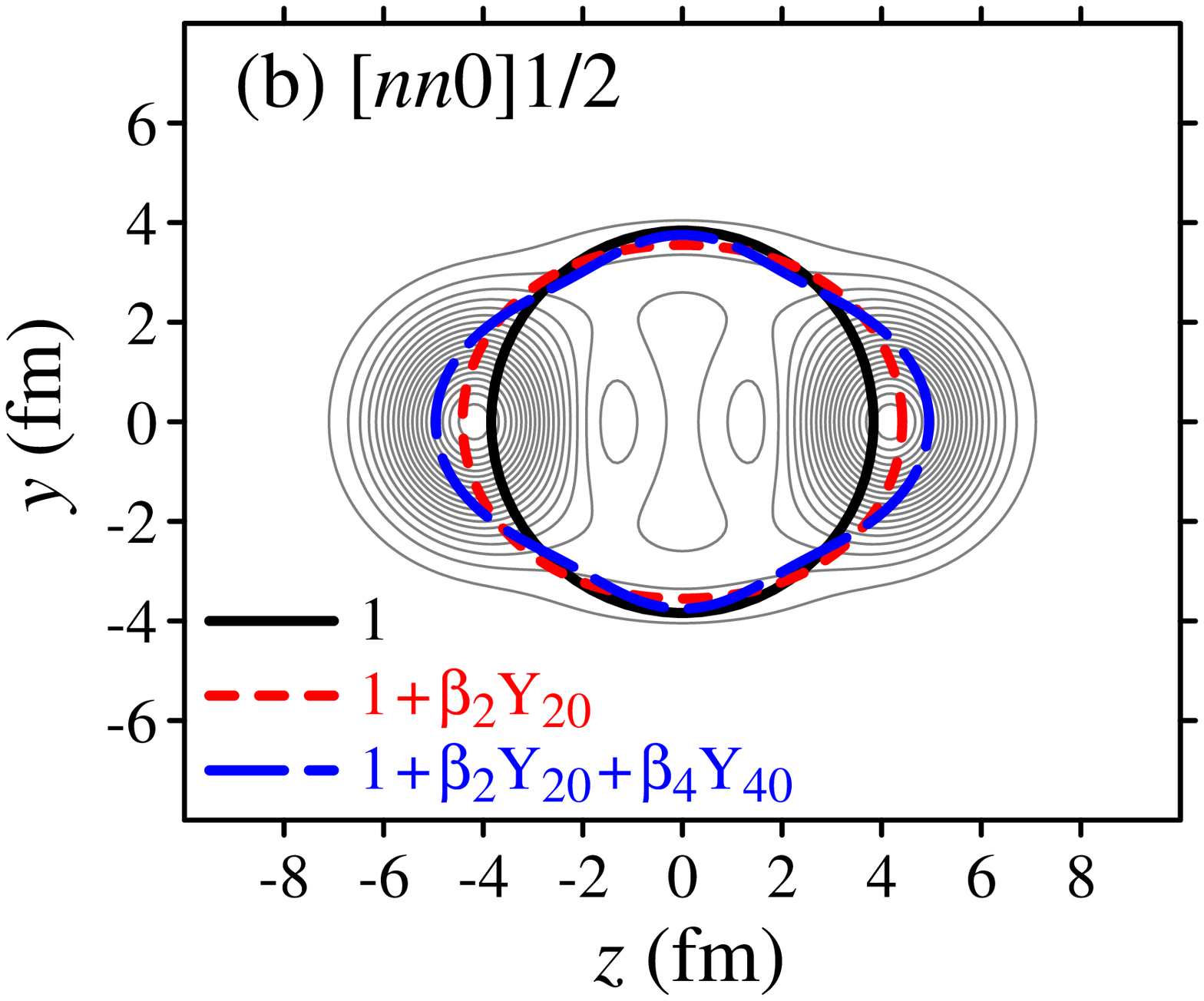,scale=0.31}
    \epsfig{file=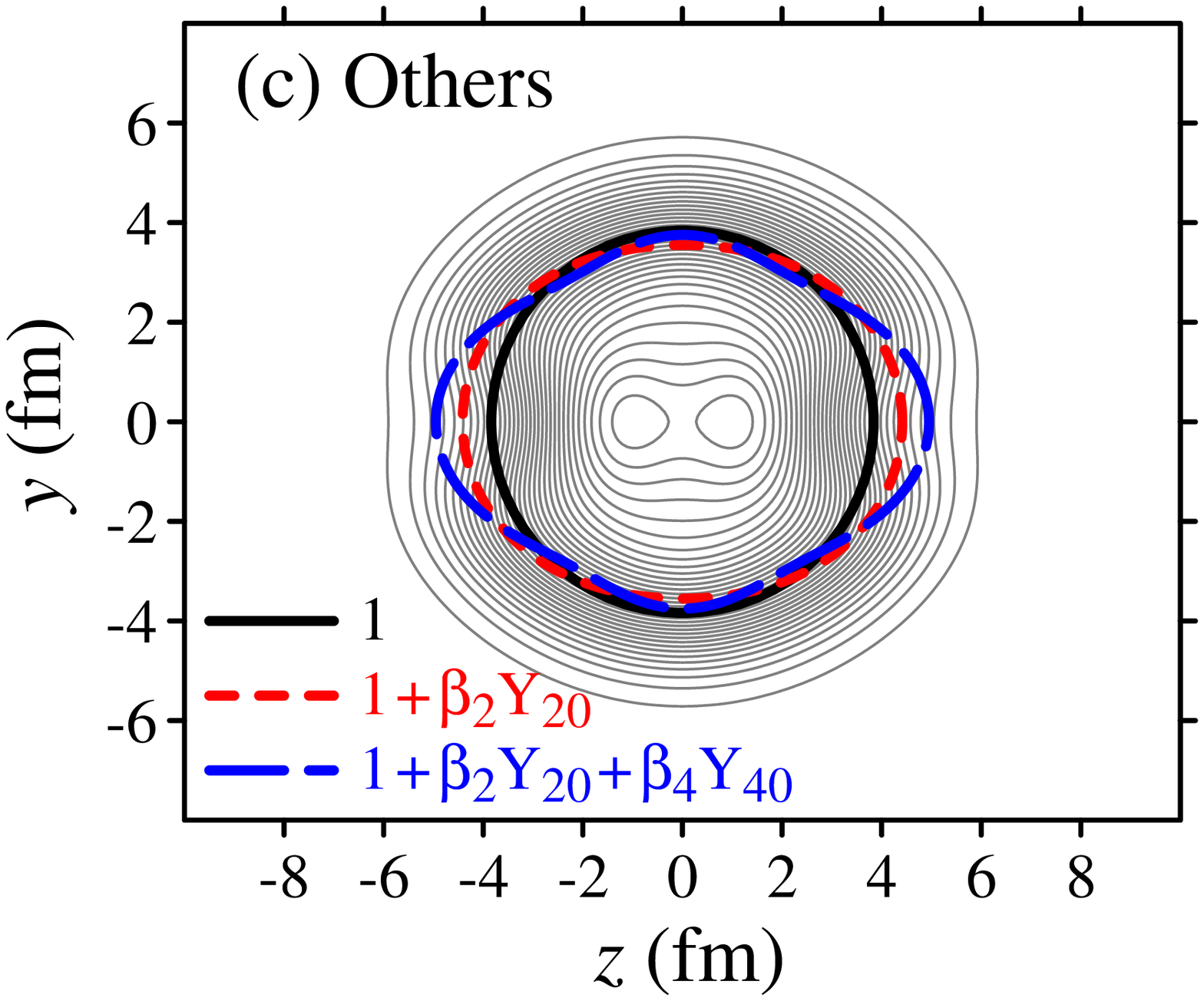,scale=0.31}        
    \caption{Contour plot of the intrinsic density distribution of $^{58}$Ti
      with the SkM$^*$ interaction,
      (a) total, (b) the sum of the
      $[220]1/2$, $[330]1/2$, and $[440]1/2$ orbits, and (c) the others.
      The sum of the density distributions of (b) and (c) corresponds
      to (a). The values at the peaks are (a) 0.184 (b) 0.096 fm$^{-3}$,
      and (c) 0.171 fm$^{-3}$. The contours are drawn
      in intervals of 0.005 fm$^{-3}$. Thick lines denote
      the nuclear radii considering spherical (1),
      quadrupole $(1+\beta_2Y_{20})$ deformation,
      and hexadecapole deformation $(1+\beta_2Y_{20}+\beta_{4}Y_{40})$.}
    \label{density58Ti.fig}
  \end{center}
\end{figure*}

To see the role of the elongated orbitals more quantitatively, 
we calculate the hexadecapole moment
$\left<Q_{40}\right>_{\rm sp}=
\left<[nn_z\Lambda]\Omega|Q_{40}|[nn_z\Lambda]\Omega\right>$
for each s.p. orbit with the asymptotic quantum number $[nn_z\Lambda]\Omega$,
and evaluate its cumulative sum
in order of the s.p. energy from lowest to highest.
We confirm that contributions of
the other hexadecapole moments $\left<Q_{4m}\right>_{\rm sp}$ with $m\neq 0$
for all contributed s.p. orbits
are negligible at most $\approx 1$ fm$^{4}$,
and thus $\left<Q_{40}\right>_{\rm sp}$ can be a good measure of
the nuclear hexadecapole deformation.
As we see in Sec.~\ref{def.sec},
  the SkM$^*$ interaction can be the most favorable choice among
  the three interactions employed in this paper.
  Hereafter we only discuss the results obtained with the SkM$^*$ interaction
otherwise noted.
Figure~\ref{cumulative.fig} compares these obtained cumulative sum
for $N=38$ isotones, where the $\beta_4$ values are
largest.
We sum up proton and neutron contributions simultaneously
and hence only the neutron contribution is considered
when the horizontal axis exceeds the proton number for each isotope.
We find sudden increases at $N=10$, 22, and 36,
which correspond to the occupation of the $[nn0]1/2$ orbital
with $n=2$, 3, and 4, respectively.
By further adding more neutrons to these neutron numbers,
the hexadecapole moment decreases due to the occupation
of less prolate orbitals.
The $\left<Q_{40}\right>_{\rm sp}$ value tends to be large
for prolately deformed orbitals and the smaller
$\left<Q_{40}\right>_{\rm sp}$ value
is found for oblate deformation, which can be expected from
the definition of $Q_{40}$ of Eq.~(\ref{q40.eq}).
An approximate expression of
  the hexadecapole moment of the axially
  symmetric anisotropic harmonic-oscillator s.p. orbit
  with the asymptotic quantum number $[nn_z\Lambda]$
  can be obtained as~\cite{BM2}
  \begin{align}
   &\left<nn_z\Lambda\right|Q_{40}\left|nn_z\Lambda\right>
   =\sqrt\frac{9}{4\pi}
    \left(\frac{\hbar^2}{2M\omega}\right)^2\notag\\
    &\times \frac{3}{4}(27n_z^2-22n_zn+3n^2-\Lambda^2-6n_z-2n),
  \end{align}
  where $M$ is the nucleon mass.
  By taking $\hbar \omega=40A^{-1/3}$ MeV with $A=62$,
  we get $\left<nn0\right|Q_{40}\left|nn0\right>=43, 130,$ and 260 fm$^4$
  for $n=2,3$, and 4, respectively, which roughly explain 
  the trend obtained by the HF calculations:
  The corresponding increases of the cumulative
  $\left<Q_{40}\right>_{\rm sp}$
  can be found in Fig.~\ref{cumulative.fig}, whose values
  are 191.9/4=48.0, 379.7/4=94.9, and 332.7/2=166.3 fm$^4$
  at the neutron/proton number 10, 24, and 36 for $^{62}$Cr.
The contribution of the [440]1/2 orbit is large and comparable to that of
the [330]1/2 orbit despite that only 2 neutrons are occupied in this orbit,
while in the [220]1/2 and [330]1/2 orbits additional 2 protons
contribute to the hexadecapole moment.

For $^{64}$Fe, an increase of the $\left<Q_{40}\right>_{\rm sp}$ value
is rather milder than the others. Since the proton number $Z=26$ fills
the [312]5/2 orbit,
in which $\left<312|Q_{40}|312\right>$ value is negative,
leading to smaller hexadecapole deformation in total.
In the case of light nuclei, this proton configuration effect is more drastic.
For example, for Ne, Mg, and Si in the island of inversion near $N=20$,
a sudden increase of the $\left<Q_{40}\right>_{\rm sp}$ value
is also found when the $[330]1/2$ orbit is occupied
at $N=20$ for $^{30}$Ne and $^{32}$Mg.
In fact, the $\beta_4$ value increases from $N=18$ to 20:
from $\beta_4=0.03$ to 0.22 for Ne,
from $\beta_4=0.00$ to 0.15 for Mg with the SkM$^*$ interaction.
In contrast, for Si, the state exhibits a spherical shape
because the proton number $Z=14$ favors the oblate deformation.
We also note that the large $\beta_4$ values for $N\approx Z$
of Ti, Cr, and Fe isotopes come from
large hexadecapole moments due to the occupancy of the $[330]1/2$ orbit
for both proton and neutron.

The density distribution of the deformed nuclear state
offers a more intuitive picture of the role of the intruder orbits for
the nuclear hexadecapole deformation.
Figure~\ref{density58Ti.fig} (a) draws a contour plot
of the intrinsic density distribution of $^{58}$Ti,
where the last two neutrons fill in the $[440]1/2$ orbit
with the SkM$^*$ interaction regarded as the edge of the island of inversion.
The nuclear radius $R$, $R(1+\beta_2Y_{20})$,
and $R(1+\beta_2Y_{20}+\beta_{4}Y_{40})$
are also plotted as a guide of the nuclear deformation.
As clearly seen in the figure,
a nuclear radius only with the quadrupole component
$(1+\beta_2Y_{20})$ does not describe the total density distribution
properly, while the nuclear radius that includes the hexadecapole component
nicely follows the contour of the total density distribution,
showing a ``lemon'' like shape.
Figure~\ref{density58Ti.fig} (b) displays
the intrinsic densities of the sum of the
$[220]1/2$, $[330]1/2$, and $[440]1/2$ orbits
which give the three largest hexadecapole moments,
and the remaining density subtracted from the total one
is also plotted in Fig.~\ref{density58Ti.fig} (c).
The role of these elongated orbitals is apparent:
The kurtosis of the total density distribution comes from
these orbitals, while the remaining density shows an almost spherical shape.

In general, the occupation of
the $[nn0]1/2$ orbitals strongly enhances the hexadecapole deformation.
A sudden increase of both the nuclear quadrupole and hexadecapole
deformation is a strong indication of the edge of the island of inversion.
Determination of the nuclear hexadecapole deformation
will have of particular importance
as it is sensitive to the occupation of the intruder $[nn0]1/2$ orbit.
We note that the inclusion of the pairing correlations
  induces the fractional occupation number of the s.p. orbits
  near the Fermi level.
  An increase of the nuclear deformation parameters
  can be somewhat milder compared to the present results.

\subsection{Quadrupole and hexadecapole deformation effects on
  nuclear radius and surface density profile}
\label{diff.sec}

\begin{figure*}[ht]
  \begin{center}
    \epsfig{file=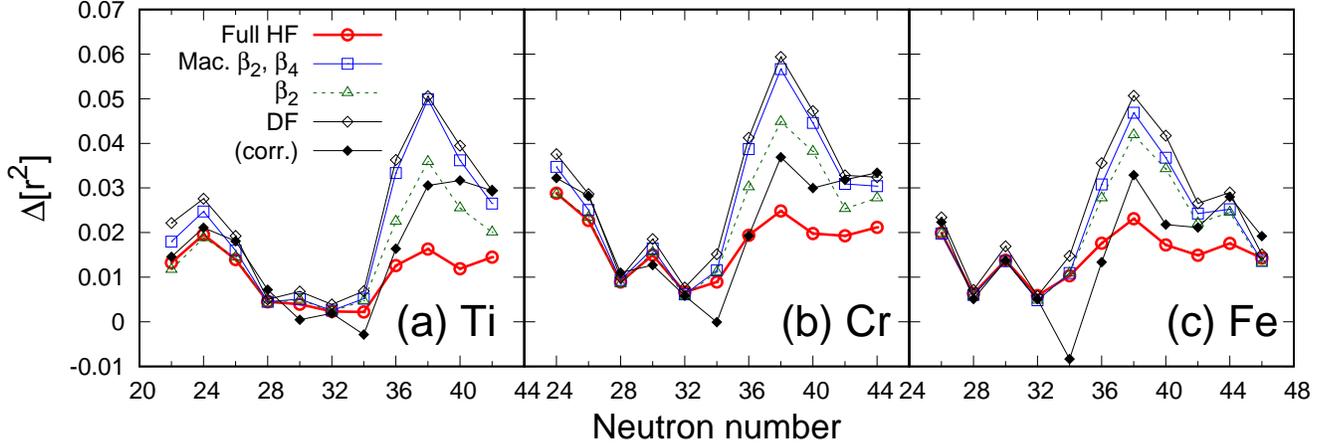, scale=1.3}
    \caption{Comparison of the relative difference of the square radii
      between the full and spherical constrained HF and the macroscopic
      formula of Eq.~(\ref{radiusformula.eq}) including
      both the quadrupole and hexadecapole deformation parameters
      ($\beta_2, \beta_4$), and that only with $\beta_2$
      for (a) Ti, (b) Cr, and (c) Fe isotopes.
      The SkM$^*$ interaction is employed for the HF calculations. 
      The results with the deformed Fermi-type (DF) density distributions
      are also plotted. See text for details.}
    \label{drcorr.fig}
  \end{center}
\end{figure*}

As we see in Sec.~\ref{hex.sec},
the strong nuclear deformation is induced by
the occupation of the intruder orbit and
crucially affects the nuclear density profile
near the nuclear surface.
This appears simply as the enhancement of the nuclear radius.
One may think that the deformation parameters can be extracted
from the change of the nuclear radius.
For small surface deformation
  with $1+\sum_{\lambda \geq 2 \in {\rm even}}\bar{\beta}_\lambda Y_{\lambda 0}$,
the following formula has often been used
to estimate the enhancement of the nuclear radius
from the spherical limit~\cite{BM, Sumi12}
\begin{align}
  \Delta[r_{\rm mac.}^2]  \approx \frac{5}{4\pi}
  \sum_{\lambda \geq 2 \in {\rm even}}\bar{\beta}_\lambda^2.
\label{radiusformula.eq}
\end{align}
We note that the hexadecapole deformation also induces
the radius enhancement, which is usually ignored.
However, we see the calculated $\beta_4$ values grow for $N> 34$
and become even comparable to the $\beta_2$ value.
Since Eq.~(\ref{radiusformula.eq}) could offer a direct relationship
between the nuclear radius and deformation,
it is worthwhile to investigate the applicable range of the formula.
To prepare the nuclear radii with a spherical limit,
we calculate the nuclear radii obtained from the spherical constrained
HF calculations using the filling approximation~\cite{Horiuchi20,Horiuchi21}.
Figure~\ref{drcorr.fig} compares
the the relative difference between the mean-square matter radii
of the full and spherical constrained HF calculations
$\Delta[r_{\rm HF}^2]=[r_{\rm m}^2{\rm (full)}-r_{\rm m}^2{\rm (sph.)}]/r_{\rm m}^2({\rm sph.})$ and $\Delta[r^2_{\rm mac.}]$ with 
  $\bar{\beta}_2, \bar{\beta}_4$ and only with $\bar{\beta}_2$.
  Here we regard $\bar{\beta}_{\lambda}$ as
  $\beta_{\lambda}$ which is calculated from the HF intrinsic moments
  of Eqs.~(\ref{beta2.eq}) and (\ref{beta4.eq}).
The results with the SkM$^*$ interaction are shown as it exhibits
the largest deformation among the other interactions.
We find that the simple formula of Eq.~(\ref{radiusformula.eq})
works well for $N\leq 34$ where $|\beta_2|\lesssim 0.2$.
The $\beta_4$ contribution is minor as the $\beta_4$ value is also small
in such small $|\beta_2|$ values.
For $N\geq 36$, the square radii from the formula
significantly overestimate the prediction of the HF calculations.
We also find large $\beta_4$ contributions
accompanied with large quadrupole deformation $\beta_2 \gtrsim 0.2$,
which induce a further deviation from the HF result.
We also evaluate the hexacontatetrapole deformation parameters
$\beta_6$ for all nuclei studied in this paper
and confirm they are small, less than $0.01$.

To incorporate the finite thickness of the nuclear surface
which is ignored in Eq.~(\ref{radiusformula.eq}),
  we consider a deformed Fermi-type (DF) density distribution
  \begin{align}
    \rho_{\rm DF}(r)=\frac{1}{4\pi}\int d\hat{\bm{r}}
    \frac{\rho_0}{1+\exp\left[(r-R(\theta))/a(\theta)\right]}
\label{dfermi.eq}
  \end{align}
  with axially symmetric deformed nuclear surface~\cite{BM2,Scamps13}
  \begin{align}
    R(\theta)&=R_0^\prime\left[1+\bar{\beta_2}Y_{20}(\theta)+\bar{\beta_4}Y_{40}(\theta)\right],\\
    a(\theta)&=a_0
    \sqrt{1+\left(\bar{\beta_2}\nabla Y_{20}(\theta)|_{r=R(\theta)}+\bar{\beta_4}\nabla Y_{40}(\theta)|_{r=R(\theta)}\right)^2}.
\label{dsurface.eq}    
    \end{align}
First, $R^\prime_0=R_0$ and $a_0$ are determined by a least-square fitting
of the spherical constrained
HF density as prescribed in Refs.~\cite{Hatakeyama18,Choudhary21}
with a spherical limit $\bar{\beta_2}, \bar{\beta_4}=0$.
The $\rho_0$ value is determined by the normalization
$4\pi \int_0^\infty r^2 \rho_{\rm DF}(r)=A$ for given $R_0$ and $a_0$.
Then $\bar{\beta_2}$ and $\bar{\beta_4}$ are determined to
simultaneously reproduce the $\beta_2$ and $\beta_4$ values
obtained from the HF calculation.
The $R^\prime_0$ value is uniquely determined by the volume conservation.
Note that $\bar{\beta_2}$ and $\bar{\beta_4}$ correlate
with the intrinsic deformation parameters $\beta_2$ and $\beta_4$,
that is, the surface deformation with
$\bar{\beta_2}$ $(\bar{\beta_4})$ also induce $\beta_4$ ($\beta_2$)
in the intrinsic density distribution defined in Eq. (\ref{dfermi.eq}).
Figure~\ref{drcorr.fig} shows these $\Delta [r^2]$
obtained with the DF density distributions.
The results are almost identical with the ones obtained by the formula
of Eq.~(\ref{radiusformula.eq}) which assumes a sharp cut radius.  
This indicates that the surface density profile
in the island of inversion cannot be explained
by simple geometric deformation, i.e,
the $R_0$ and $a_0$ values are no longer fixed parameters.

\begin{figure*}[ht]
  \begin{center}
    \epsfig{file=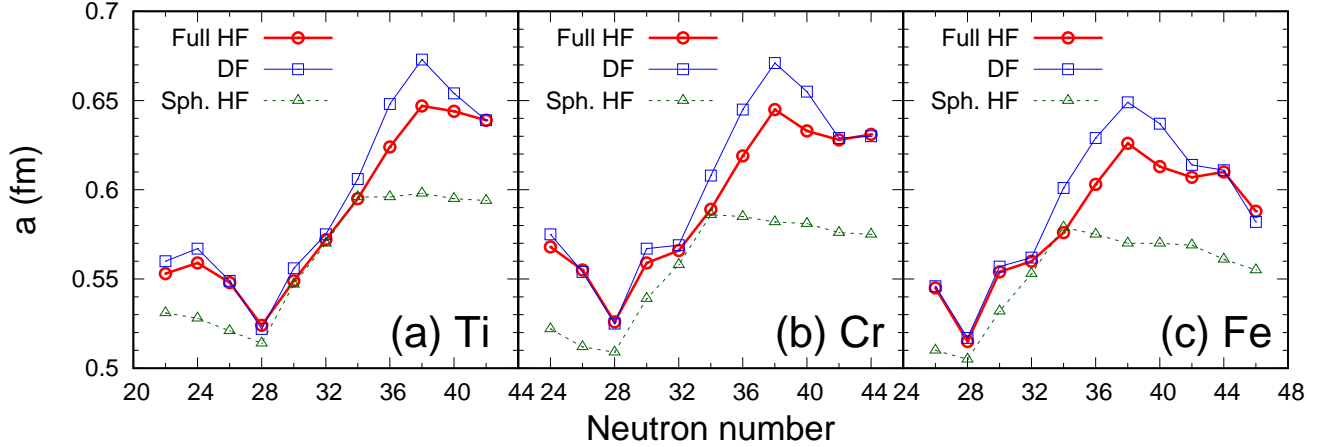, scale=1.3}
    \caption{Diffuseness parameter extracted from
      the full HF, deformed Fermi-type (DF),
      and spherical constrained HF densities
      for (a) Ti, (b) Cr, and (c) Fe isotopes.
      The SkM$^*$ interaction is employed. See text for details.}
    \label{diff.fig}
  \end{center}
\end{figure*}
 
To quantify the changes of the surface density by the nuclear deformation,
we evaluate the resultant nuclear surface diffuseness $a$
in a spherical limit
for each density distribution by using the same way described above.
Note that the one with the spherical constrained HF corresponds to $a_0$.
Figure~\ref{diff.fig} plots the diffuseness parameters 
of the full HF, DF, and spherical constrained HF.
The behavior of the $a_0$ value can be explained by
considering a standard shell model filling~\cite{Horiuchi21b}:
The constant behavior is due to
the occupation of the high-$l$ orbits, $0f_{7/2}$, $0f_{5/2}$, and $0g_{9/2}$,
while the low-$l$ orbits, $1p_{3/2}$ and $1p_{1/2}$ in $28\leq N\leq 34$
enhances the surface diffuseness.    
Compared with the diffuseness parameter extracted from the spherical constraint
HF density, the $a$ value of both HF and DF significantly
enhanced by the nuclear deformation.
The diffuseness parameters obtained by the DF distributions
overestimate the full HF ones in $36\leq N \leq 40$
despite that the DF results nicely agree with the HF results
of the outside of the island of inversion.
We calculate the $\Delta[r^2_{\rm m}]$ value
with the DF density by varying $a_0$ so as to reproduce 
the resultant diffuseness parameter $a$ for the HF calculation.  
As shown in Fig.~\ref{drcorr.fig}, the results are improved
for $36 \leq N\leq 40$, implying the surface diffuseness changes
beyond the geometrical one, while the
systematic behavior due to the nuclear deformation
is still present, and thus the nuclear deformation parameters
can be constrained if one knows details about the nuclear density
profile near the nuclear surface.

We note that this reduction of the diffuseness parameter
is non-trivial because the nuclear deformation
correction to the surface diffuseness is always positive.
See Eq.~(\ref{dsurface.eq}).
We remark that anti-correlation between the nuclear deformation
and surface diffuseness was reported in Ref.~\cite{Scamps13}.
Reference~\cite{Horiuchi21} demonstrated that
the nuclear deformation changes the density profile drastically
and depends strongly on the shell structure near the Fermi level.
From a microscopic point of view,
in $N>34$, the intruder [440]1/2 orbit induces
the mixing of the spherical $0g_{9/2}$, $1d_{5/2}$, and $2s_{1/2}$ orbits.
The lower angular momentum orbits, $1d_{5/2}$ and $2s_{1/2}$,
play a role to enhance the nuclear radius owing
to the large penetrability of the neutrons near the
nuclear surface, while the $0g_{9/2}$ orbit 
suppresses the enhancement of the nuclear radius,
which has the sharpest nuclear surface distribution.
In fact, the neutron occupation number of the $0g_{9/2}$, $1d_{5/2}$ and
$2s_{1/2}$ orbits was found to be 2.61, 0.95, and 0.04, respectively
for $^{62}$Cr~\cite{Horiuchi21}. We also remark that this large mixing
of the $0g_{9/2}$ orbit is consistent with
the finding of Ref.~\cite{Lenzi10}.
Consequently, resulting small surface diffuseness leads to the reduction
of $\Delta[r^2_{\rm m}]$ at $N\geq 36$.
This shell effect induces the non-trivial change of the nuclear
density profile, which cannot be explained from the simple geometric formulae.
Similar behavior but a mixing of $1p_{3/2}$ and $0f_{7/2}$ at
the island of inversion $N \gtrsim 20$ for Ne and Mg isotopes
was reported in Ref.~\cite{Choudhary21}.  

\begin{figure*}[ht]
  \begin{center}
    \epsfig{file=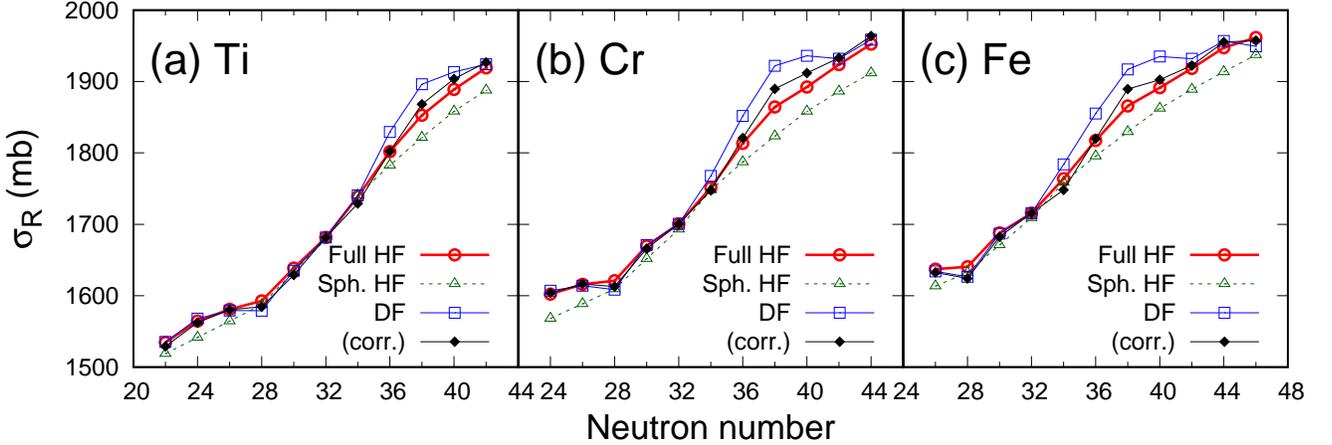, scale=1.3}
    \caption{Total reaction cross sections
      for (a) Ti, (b) Cr, and (c) Fe isotopes
      calculated with the full HF, spherical HF,
      deformed Fermi-type (DF),
      and diffuseness-corrected DF density distributions.
      Note that the cross sections with the DF density distributions
      are normalized to the full HF result at $N=34$.
      The SkM$^*$ interaction is employed for the HF calculations.}
    \label{rcs_diff.fig}
  \end{center}
\end{figure*}

Figure~\ref{rcs_diff.fig}
displays the total reaction cross sections calculated
with the full HF, spherical HF, DF, diffuseness-corrected DF
density distributions of Cr isotopes at 240 MeV/nucleon.
For the sake of comparison,
the cross section values with the DF density distributions
are normalized to the full HF result at $N=32$.
The use of a carbon target has an advantage
for studying details of the nuclear density profile, i.e.,
deformation effect on the nuclear surface
because the density profile near the nuclear surface
is more pronounced than a proton target~\cite{Horiuchi14}.
In fact, we see a strong sensitivity
in the cross section differences in $N \gtrsim 34$,
which can be distinguished
with the present experimental precision~\cite{Tanaka20,Bagchi20}.
Therefore, a systematic measurement of
  the total reaction cross sections in the isotope chains
  is strongly desired to get the structure information
at the edge of and in the island of inversion.
We also remark that proton-nucleus elastic scattering
at forward angles is useful to obtain the nuclear surface
diffuseness~\cite{Hatakeyama18, Choudhary20,Choudhary21}
as complementary evidence.

It should be noted that this finding
opens the possibility of determining the hexadecapole
deformation parameter from measurements of the total reaction cross sections.
For example, one may assume the DF density distribution
of Eq.~(\ref{dfermi.eq}), which includes four free parameters.
They can be fixed by measurements of
the total reaction cross sections at various
incident energies and target nuclei.
The parameters can further be constrained
with information of the quadrupole deformation which can be
deduced from a measurement of the electric-quadrupole transition strength.
A careful investigation is necessary to extract these parameters quantitatively.

\section{Conclusion}
\label{Conclusions.sec}

We have made a systematic analysis of nuclear deformation and
discuss their effects on the nuclear radius or
the total reaction cross section for even-even neutron-rich
Ti, Cr, and Fe isotopes using reliable microscopic
mean-field structure and reaction models. 
We have evaluated three standard sets of effective interactions
for the mean-field calculations
to examine the model dependence that comes from the nuclear deformation.

Using those obtained density distributions,
we have calculated the total reaction cross sections
without introducing any free parameter.
In general, the cross section is enhanced
if the nucleus exhibits strong deformation.
We show that the enhancement is significant
and can be identified with the recent experimental precision.
Given the present comparison of the theoretical calculations
with the available experimental 
evaluations of the quadrupole deformation and the recent data
of the two-neutron separation energies,
 $N>36$ is most likely in the island of inversion,
where a sudden increase of nuclear deformation is predicted.
The total reaction cross section offers more
concrete evidence
to determine the edge of the island of inversion near $N=40$.

Characteristic nuclear deformation is found in the island of inversion:
Strong hexadecapole deformation occurs simultaneously with
the quadrupole deformation
due to the occupation of the strongly deformed $[nn0]1/2$ Nillson orbit.
This characteristic structure
drastically changes the density profile of
these nuclei. However, we find that the nuclear radius enhancement in
the island of inversion cannot be explained
by a simple geometrical deformation model, which implies
non-trivial changes of the density profile
that come from the shell structure near the Fermi level.
This motivates us to study higher-order size properties of nuclei
further than the nuclear radius, e.g.,
nuclear diffuseness~\cite{Hatakeyama18},
and higher radial moments~\cite{Kurasawa19}.

The determination of both the nuclear radii and surface
density profile will open a way to determine
the deformation parameters of these nuclei, especially for
the hexadecapole deformation.
For this purpose, measurements of
the total reaction cross sections for those isotopes at different incident
energies and target nuclei are highly desired.
Complementary information of experimental quadrupole deformation
parameter and nuclear surface diffuseness can further help
determine the higher-order term of the nuclear deformation.

\acknowledgments
This work was in part supported by JSPS KAKENHI Grants No.\ 18K03635.
We acknowledge the collaborative research program 2021, 
Information Initiative Center, Hokkaido University.

\end{document}